# Scalable Miniature On-chip Fourier Transform Spectrometer
# For Raman Spectroscopy

Sarp Kerman[1,2,†], Xiao Luo[2], Zuoqin Ding[2,] Zhewei Zhang[3], Zhuo Deng[3], Xiaofei Qin[2], Yuran Xu[2,4], Shuhua Zhai[2], Chang Chen[1,2,†]

[1]*Institute of Medical Chip, Ruijin Hospital, Shanghai Jiao Tong University School of Medicine, Shanghai, China*
[2]*Shanghai Photonic View Technology Co., Ltd., Shanghai, China*
[3]*Shanghai Industrial µTechnology Research Institute (SITRI), Shanghai, China*
[4]*Friedrich Schiller University Jena, Jena, Germany*
[†]Email: sarp.kerman@photonicview.com, changchen@sjtu.edu.cn

Miniaturized spectrometers for Raman spectroscopy have the potential to open up a new chapter in sensing. Raman spectroscopy is essential for material characterization and biomedical diagnostics, however, its weak signal and the need for sub-nanometer resolution pose challenges. Conventional spectrometers, with footprints proportional to optical throughput and resolution, are difficult to integrate into compact devices such as wearables. Waveguide-based Fourier Transform Spectrometers (FTS) enable compact spectrometers, and multi-aperture designs can achieve high throughput for applications such as Raman spectroscopy, however, experimental research in this domain remains limited. In this work, we present a multi-aperture SiN waveguide-based FTS overcoming these limitations and enabling Raman spectroscopy of isopropyl alcohol, glucose, Paracetamol, and Ibuprofen with enhanced throughput. Our spectrometer chip, fabricated on a 200 mm SiN wafer, with 160 edge-coupled waveguide apertures connected to an array of ultra-compact interferometers and a small footprint of just 1.6 mm x 4.8 mm, achieves a spectral range of 40 nm and a resolution of 0.5 nm. Experimental results demonstrate that least absolute shrinkage and selection operator (LASSO) regression significantly enhances Raman spectrum reconstruction. Our work on waveguide-based spectrometry paves the way for integrating accurate and compact Raman sensors into consumer electronics and space exploration instruments.

Optical spectrometers play a crucial role in a wide range of applications such as laser and material characterization, process monitoring, biomedical research, forensic analysis, and space applications[1–6]. Conventional spectrometers based on diffractive elements face a trade-off between resolution and optical throughput[7]. Fourier transform spectrometers (FTS) overcome this limit by eliminating the need for slits[8] (Jacquinot's advantage). Common FTS instruments consist of a moving mirror as part of a Michelson interferometer and reconstruct the spectrum in the Fourier domain by capturing the interferometer response with varying frequencies over wavelength[9,10]. Typical stationary spatial heterodyne FTS designs split light and combine after reflection from tilted surfaces, to simultaneously image the response of multiple interferometers onto an array of photon detectors[11–15]. Both approaches benefit from noise reduction thanks to multiple measurements[16] (Fellgett's advantage).

A novel waveguide-based spatial heterodyne FTS was presented as a path to miniaturize optical spectrometers in 2007 by Florjanczyk et al.[17]. This structure consisted of a number ( $N$ ) of asymmetric Mach-Zehnder Interferometers (MZI) with progressively increasing path length differences ($\Delta L$) between the arms. The spectral resolution ($\delta_\lambda$) is determined by the free spectral range (FSR) of the MZI with the longest $\Delta L$ , while the operational range is defined by the FSR of the entire array as $N\delta_\lambda/2$ . A great deal of research[18–20] has been devoted to enhancing the range and resolution[21–31] and decreasing the footprint primarily using actively controlled interferometers and resonators[24–31]. However, a limited amount of research has been performed on increasing the optical throughput[17,32–36], which is vital for detecting a weak and widespread signal such as Raman scattering.

Raman spectroscopy stands as an important tool for material characterization. Combining waveguide-based FTS and Raman spectroscopy would bring a valuable and affordable tool for next-generation material characterization. While waveguide-based FTS can achieve sub-nanometer resolution, Raman spectroscopy requires high optical throughput due to its broad angular range, particularly in measuring scattering substances. Additionally, the excitation spot may be large in case a high-power multimode laser is used or the power may be intentionally spread across a broader area to avoid damaging the substance. Both the angular range and the excitation area determine the optical throughput as $G \approx n^2 S\Omega = \pi S. NA^2$ , where $n$ is the refractive index, $S$ is the area, $\Omega$ is the solid angle and NA is the numerical aperture. Commercial Raman spectrometers typically offer comparable or an order of magnitude less throughput than a multimode fiber (MMF) with 105 µm core diameter and NA = 0.22. In contrast, a single-mode waveguide operating at 850 nm wavelength has about 1800 times less throughput. Multi-aperture waveguide-based FTS can provide high optical throughput and spectral resolution by using multiple interferometers with independent inputs.

The choice of waveguide material is crucial for combining Raman spectroscopy with waveguide-based

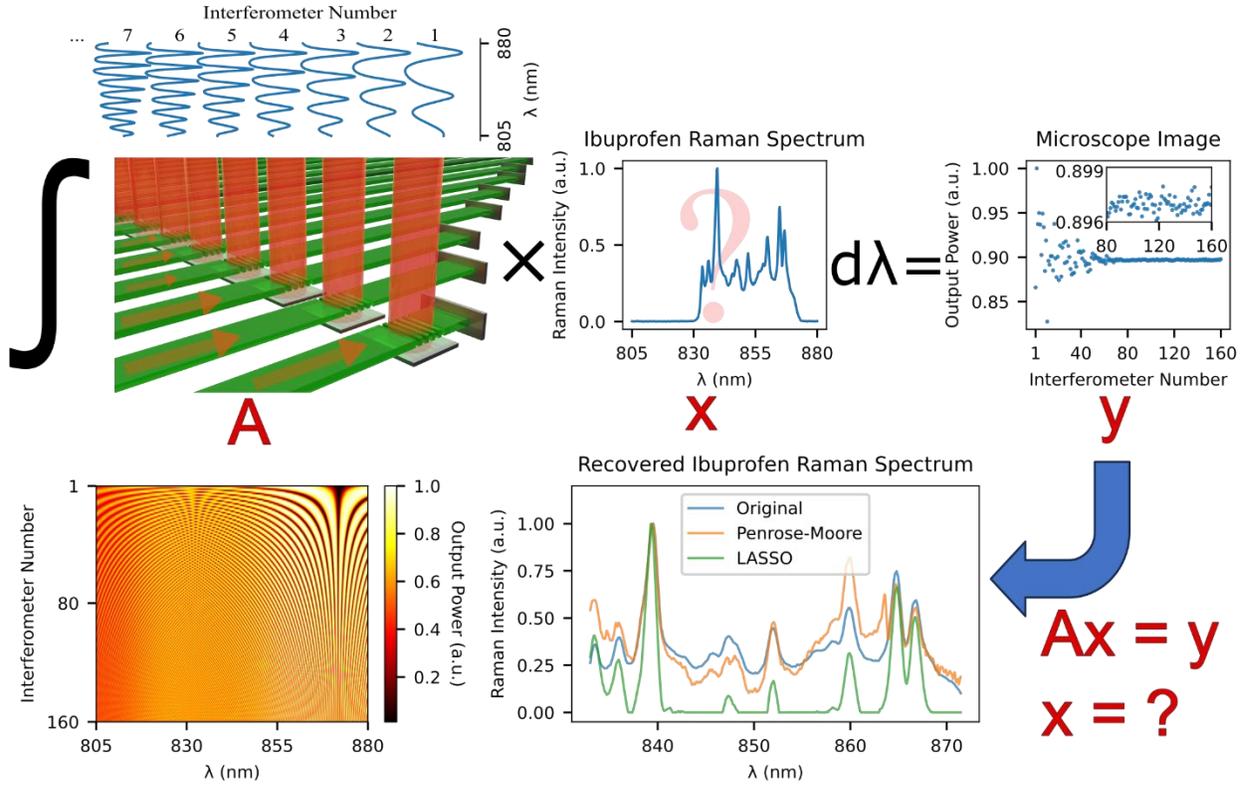

**Figure 1. Working principle of the miniature Raman spectrometer.** The multi-aperture waveguide-based FTS (*A*) collects an exemplary Raman spectrum with high optical throughput. The output power from each interferometer (*y*) is measured, and the Raman spectrum (*x*) is reconstructed by solving the linear matrix equation ($Ax = y$), where $A$ represents the transform matrix of the FTS.

FTS, as Raman spectroscopy usually operates in the visible and near-infrared (NIR) range of the spectrum. For instance, with a common Raman excitation source at 785 nm, the most relevant spectral region spans 800-950 nm. Although most waveguide-based FTS was demonstrated on Si photonics, Si is opaque at these wavelengths. Silicon's high refractive index contrast enables dense stacking of a large number of waveguides without any significant crosstalk. Visible transparent platforms such as silica or polymer waveguides require tens of micrometers spacing between waveguides, limiting the number of interferometers to thousands per reticle. SiN has emerged as a matured waveguide material with transparency in the visible-NIR spectrum and a relatively high refractive index[37] ($n \sim 2$). SiN allows waveguide separation of a few micrometers, enabling reticles with tens of thousands of interferometers.

It should be noted that the estimation of the number of interferometers per reticle assumes that each interferometer is as wide as a waveguide itself. However, most interferometers, such as asymmetric MZIs, require 90° bends which substantially increases their footprint. While compact spectrometers based on stationary-wave integrated FTS (SWIFTS) have been reported[38-42], they require high-resolution imaging, active control, or a.

rigorous fabrication process, adding to the integration complexity.

In this work, we report the Raman spectroscopy detection of four different substances using a SiN waveguide-based FTS, fabricated on a 200 mm wafer using Complementary Metal-Oxide-Semiconductor (CMOS) fabrication compatible techniques, with 160 edge-coupled input apertures, a spectral range of 40 nm, and a resolution of 0.5 nm. Fig. 1 illustrates the operating principles of our spectrometer. The Raman signal, filtered to match the FTS's spectral range, was guided into individual interferometers at the edge of the chip using a cylindrical lens. The interferometers exert harmonic response with gradually increasing frequency over wavelength. Each interferometer consists of a single line of waveguide, making it very compact compared to the alternatives with on-chip branching.

The interferometer structure is similar to a SWIFTS in Lippmann configuration[42], but instead of imaging the standing wave pattern, it relies on imaging the output power of each interferometer which is wavelength dependent. A vertical grating coupler (VGC) is followed by a waveguide, which is terminated by an Al mirror. This structure allows forward and backward propagating waves to interfere, generating a beating pattern over the VGC. The position of this pattern shifts depending on the wavelength due to the cavity's phase accumulation, making the output power wavelength dependent. The output power of each interferometer (y) was captured using a microscope camera. Subsequently, the Raman signal spectrum (x) was reconstructed by solving the linear matrix equation $Ax = y$ for $x$, where $A$ represents the pre-characterized transform matrix (T-matrix) of the FTS.

An example of spectrum reconstruction is shown in Figure 1, using the Raman spectrum of Ibuprofen as the sample input signal. The transform matrix was generated from a numerical model that represents our spectrometer's behavior. As seen in the example, the output power variation between neighboring interferometers decreases rapidly for those with shorter FSRs. This makes high-resolution contributions of the FTS more prone to noise, creating a need for a robust regression method to enhance the accuracy of the reconstructed spectrum. Given the inherently weak Raman signal, a regression method that enhances the major Raman peaks and lowers the noise is essential.

We compared two regression techniques for the spectrum reconstruction: the Penrose-Moore pseudoinverse and least absolute shrinkage and selection operator (LASSO). While LASSO regression successfully provided a solution with major peaks, it suppressed minor peaks along with the noise. Moreover, LASSO fragmented the major peaks into multiple narrower peaks with small gaps between them, a profile previously reported in the literature as a poor match[29]. To mitigate this, we applied Savitzky-Golay smoothing after LASSO regression, which improved the matching with the expected Raman spectrum. Furthermore, we investigated other regression methods to evaluate their performance in improving the accuracy and reliability of the reconstructed spectrum.

**Compact Fourier Transform Spectrometer Design**

The initial step in designing the waveguide-based FTS was determining the spectrometer's center wavelength and range. We targeted a center wavelength of 860 nm with a spectral range of 40 nm, as the substances under testing exhibit distinct Raman spectra within this range for a 785 nm excitation wavelength. Given the desired resolution of 0.5 nm, we designed a spectrometer consisting of 160 interferometers.

Figure 2a presents a simplified layout of the waveguide-based FTS. Each interferometer consists of an edge coupler ( $w_E = 20.0 \ \mu m$ ), a single-mode waveguide ( $w_{WG} = 466 \ nm$ ) to filter out higher-order modes, a fully etched VGC, and an aluminum mirror on an expanded waveguide. We used $t_{SiN} = 150 \ nm$ thick low-pressure chemical vapor deposition (LPCVD) SiN platform to mitigate both edge-coupling and propagation losses. The VGC and cavity width were set as 8 $\mu m$ to control light divergence, allowing efficient power collection by the microscope system with NA = 0.14.

The interferometer design began by determining the longest cavity length and cavity length difference. The longest cavity length was set to 384.565 $\mu m$, based on the required resolution ( $\delta_\lambda = 0.5 \ nm$ ). Consequently, the cavity length difference between neighboring waveguides ( $\Delta L = 2.390 \ \mu m$ ) was determined using the Littrow wavelengths of 800 nm and 880 nm, where all interferometers give identical response, making the spectral range of the FTS between 840 nm to 880 nm. It was assumed that the Littrow wavelength at 880 nm would blue-shift due to longer propagation in the VGC, falling within the T-matrix characterization range.

The next step involved designing the VGC. The cross-section of the interferometer cavity is depicted in Figure 2b. The grating parameters, including the number of grating lines, pitch ( $p = 567 \ nm$ ) and etch width ( $e = 318 \ nm$ ), were optimized through a series of finite-difference time-domain (FDTD) simulations, using the shortest interferometer ( $L_0 = 4.555 \ \mu m$ ) as the basis. The optimization aimed to maximize both the output power and the modulation depth. Of these parameters, the number of grating lines has the greatest impact on output power, while all parameters influence the modulation depth.

Forward and backward propagating waves form a standing wave over the VGC where spatial matching of the constructive interference with a perturbation (grating etch) maximizes the outcoupling. For optimal modulation depth, all perturbation positions must align in phase with the standing wave, implying that the VGC must couple at the normal angle, effectively functioning as a distributed Bragg reflector (DBR). Back-reflection bandwidth is limited as the pitch corresponds to a second-order DBR. To optimize the output power, the back-reflected power must be minimized, which was achieved by adjusting the etch width.

To further enhance the output power, we placed an aluminum reflector beneath the VGC. The reflector distance ( $h = 325 \ nm$ ) was designed for the normal angle, to provide a sufficient bandwidth for the intended spectral range. This distance was kept as small as possible, limited by the propagation loss, to minimize performance variations due to fabrication tolerances. The top-oxide thickness ( $t_{TOX} = 3325 \ nm$ ) was determined to optimize edge-coupler efficiency and output power. The bottom oxide thickness ( $t_{BOX} = 4525 \ nm$ ) was kept large enough to isolate the silicon substrate from the waveguide.

A 200 mm wafer based on this design, along with characterization structures, was fabricated at the United Microelectronics Center (CUMEC). Figure 2c shows an image of the fabricated wafer and a spectrometer chip with dimensions of 1.6 mm x 4.8 mm. The fabrication process was verified using multiple scanning electron micrographs (SEM). Figure 2d represents an SEM image of the cross-section of one of the interferometers near its mirror. The measured waveguide thickness was close to the target of 150 nm, although this measurement's accuracy highly depends on the cleaving angle of the chip. Moreover, we observed that the mirror was inclined by 6°, which could introduce extra losses upon reflection.

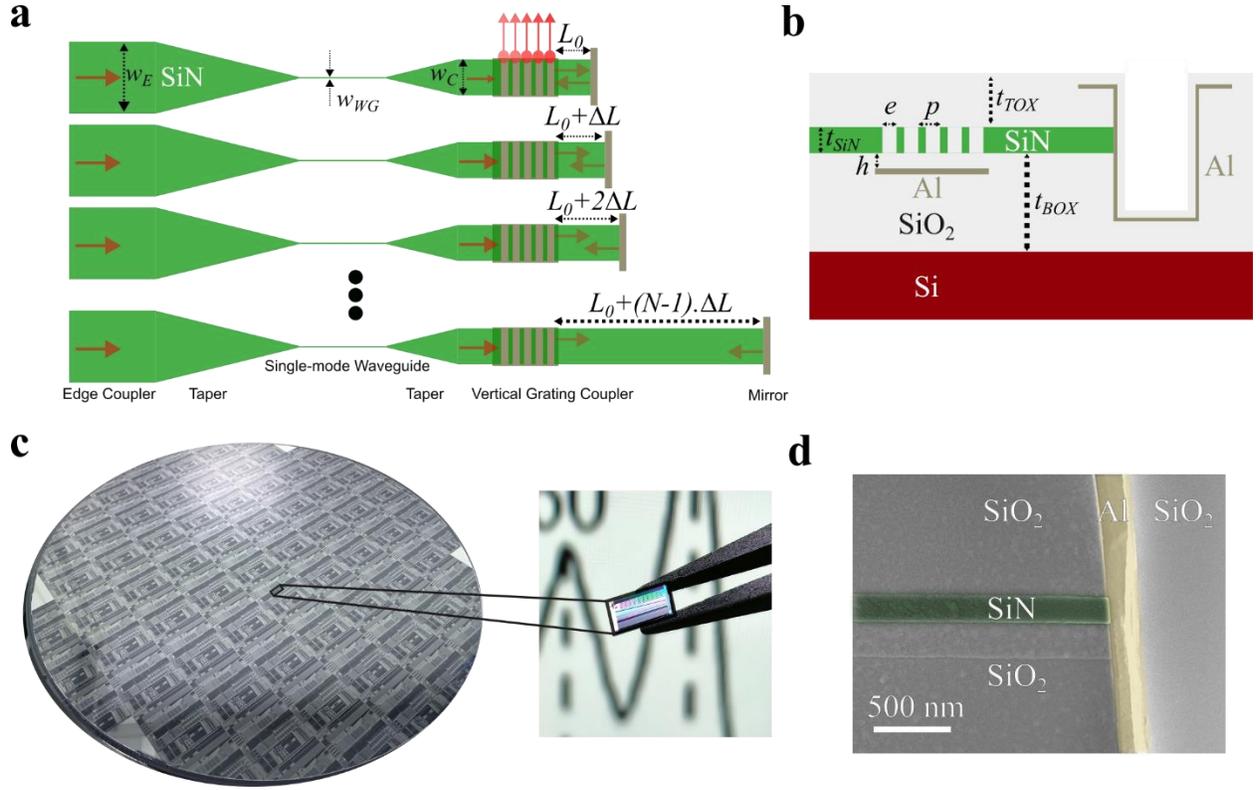

Figure 2. **The multi-aperture waveguide-based Fourier transform spectrometer. a**. Schematic top-view of edge-coupled multi-aperture waveguide-based Fourier transform spectrometer (FTS), illustrating the operation of the ultra-compact interferometers. **b**. Cross-sectional schematic of an interferometer, showing its back-reflector enhanced vertical grating coupler (VGC) and termination mirror. **c**. Pictures of the 200 mm SiN photonics wafer and one of the FTS chips, with a Raman spectrum in the background. **d**. Scanning electron micrograph of the cross-section of an interferometer, near its mirror termination.

To enhance Raman signal collection and to use TM polarization for light sheet amplitude calibration, we integrated a polarization rotating beam splitter (PRBS) into the layout. However, due to fabrication challenges, the performance for $TM_0$ to $TE_0$ coupling was suboptimal, while the $TE_0$ to $TE_0$ port remained unaffected.

## Characterization of the Fourier Transform Spectrometer

Our FTS was characterized for its wavelength response by a tunable laser source (with line-width 0.06 nm) from 805 nm to 880 nm with 0.05 nm steps. The laser light was focused into a light sheet, with a full-width half-maximum (FWHM) width of 600 µm and length of 2.9 mm, using a series of cylindrical lenses, and it was carefully aligned to the edge of the chip while monitoring the output from the VGC using a camera. Figure 3a shows a camera image of the output signal, illustrating the power variation across the interferometers for an 850 nm input. As expected, the signal was heavily modulated over the interferometers at this wavelength. A slowly varying Gaussian profile, consistent with the shape of the light sheet, was observed.

We recorded images at each wavelength within the test range and normalized the output power from each image using a Gaussian fit and the corresponding laser power. Figure 3b displays the response of four interferometers with varying lengths, showing their respective wavelength responses. We observe modulation depths ranging from 2.5 dB to 9.5 dB. The experimental results showed good agreement with our numerical model, where we used a genetic algorithm to match by varying the effective refractive index ($n_{eff}$), $h$, $t_{TOX}$ and interferometer length. A correction factor ($L_{err}$) was applied to account for potential deviations in interferometer length due to mask misalignments. The optimized parameters were: $h = 299\ nm$, $t_{TOX} = 3038\ nm$ and $L_{err} = -175\ nm$, indicating a slight shift of the mirror towards the VGC.

The mean FSR for each interferometer was calculated by applying a Fourier transform to their respective output power signals. As shown in Figure 3c, the FSR exhibits a clear trend of decreasing from 37.53 nm to 0.51 nm (which defines $\delta_\lambda$). The low variation

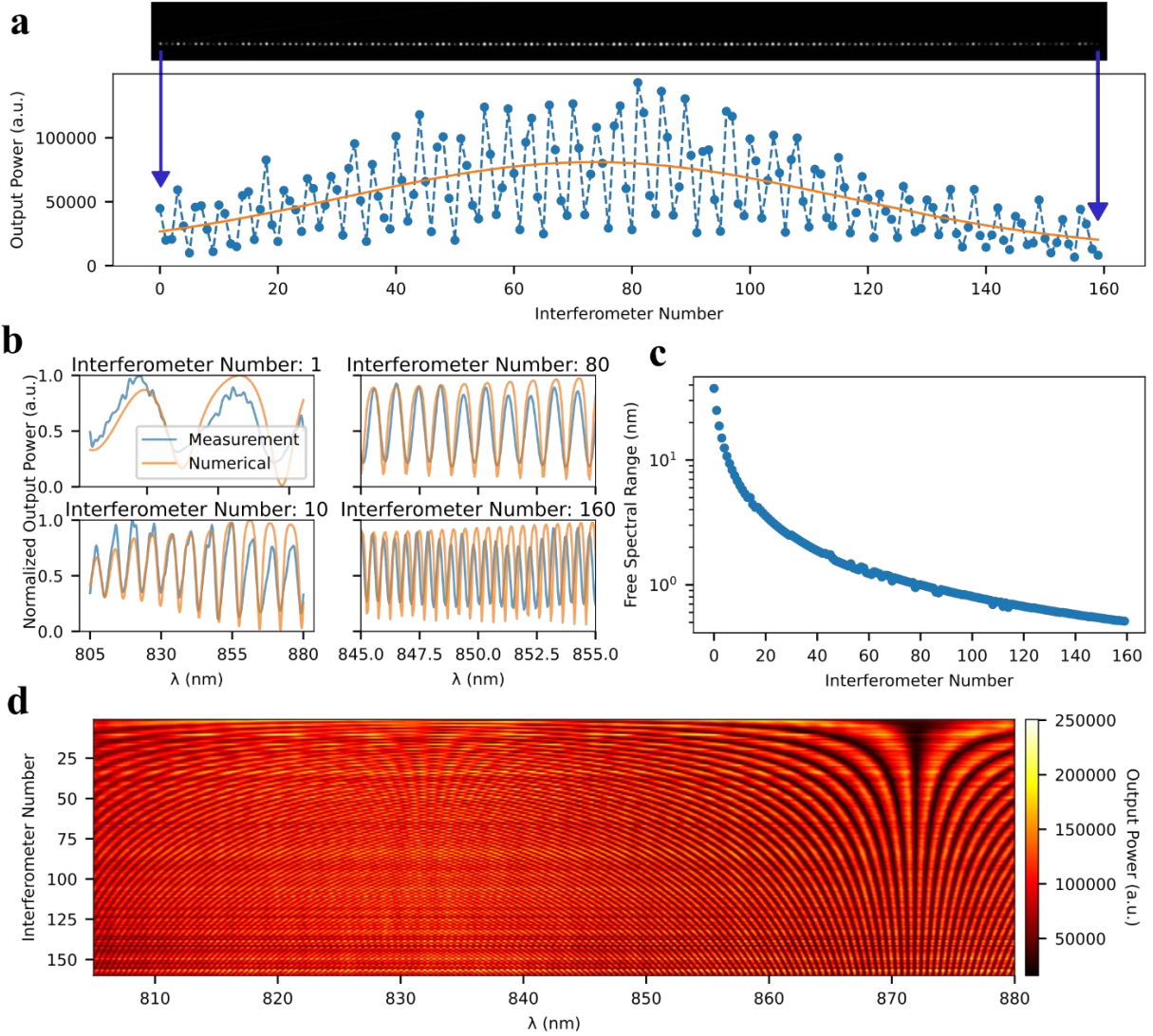

Figure 3. **Characterization results of the multi-aperture waveguide-based Fourier transform spectrometer. a**. Camera image of the vertical grating coupler's outputs and the corresponding output power of the interferometers. **b**. Measured and numerically calculated interferometer responses versus wavelength, both normalized by their respective maximum values between 805 nm and 880 nm. **c**. Free spectral range of each interferometer, averaged over the wavelength range. **d**. Transform matrix of the Fourier transform spectrometer, showing the response of all interferometers within the characterized wavelength range.

from the trend indicates low phase noise.

After normalizing the output powers of each interferometer for both the laser power and the light sheet, we compiled the data into a transform matrix ($A$), as illustrated in Figure 3d. The matrix then was used for reconstructing the laser and Raman spectrums. We observed a Littrow wavelength at $871.85\ nm$ and a wavelength where the consecutive interferometer responses phase shift by $\pi$ (called a-Littrow for convenience) at $832.10\ nm$, corresponding to a spectral range of $39.75\ nm$.

The slight discrepancies between the measured range and resolution compared to the initial design specifications can be attributed to the actual SiN waveguide thickness of $t_{SiN} = {\sim}136\ nm$, rather than the intended $150\ nm$. The thickness was estimated through simulations to match the calculated (using $\Delta L$) effective refractive indices ($n_{eff}$) for the measured Littrow and a-Littrow wavelengths, yielding values of 1.642 and 1.654, respectively. In our numerical model, illustrated in Figure 3b, the $n_{eff}$ was varied around these values for the corresponding wavelength.

**Laser Spectrum Reconstruction**

We evaluated the spectrum reconstruction performance of the FTS by repeating the T-matrix measurement. Figure 4a shows the results obtained using both Penrose-Moore pseudoinverse and LASSO regression methods. For the pseudoinverse, each peak generated a mirrored peak symmetrical around the

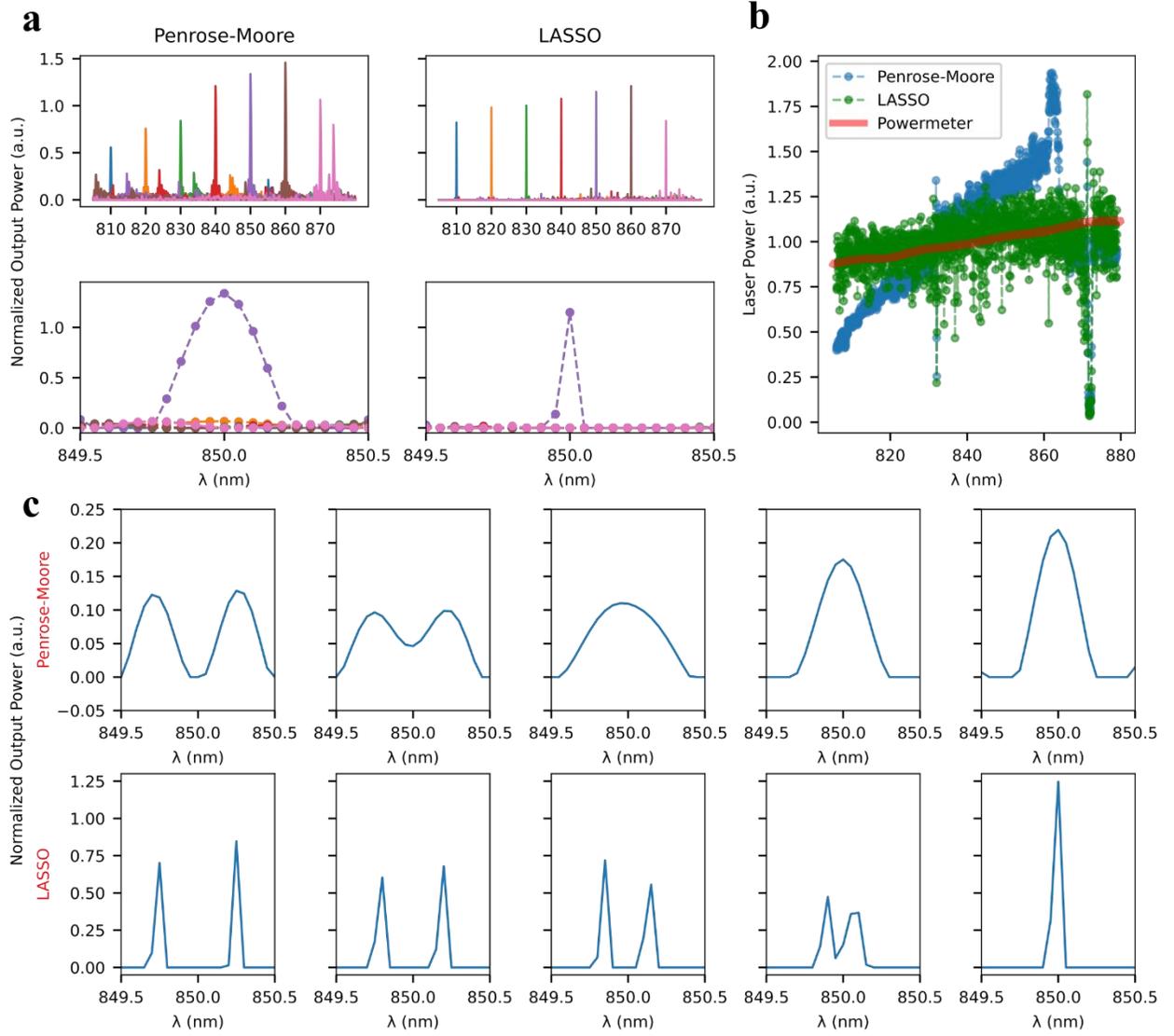

Figure 4. **Measurement results of the spectrum reconstruction of a tunable laser source. a.** Reconstructed spectra of the laser as it was tuned from 810 nm to 870 nm in 10 nm increments, using Penrose-Moore pseudoinverse and LASSO regression methods. **b.** Amplitude variation of the reconstructed peak compared to the power meter read-outs. **c.** Reconstructed spectra of the sum of output powers from two laser measurements, centered at 850 nm and separated by 0.5 nm to 0.1 nm, with 0.1 nm steps (left to right columns). The upper row shows reconstructions by Penrose-Moore pseudoinverse, while the lower row shows reconstruction using LASSO regression.

Littrow and a-Littrow wavelengths. However, these mirrored peaks exhibited smaller amplitudes due to the asymmetry of the interferometer responses. In contrast, LASSO regression effectively suppressed these mirrored peaks and narrowed the main peak, providing a broader operational range and higher resolution than expected. This is due to LASSO regression optimizing the sparsity and enhancing the peak with the best matching solution while minimizing the rest of the spectral coefficients, thus prohibiting the appearance of any artificial peaks.

To characterize the resolution, we summed the interferometer output powers for two closely spaced wavelengths with separations varying from 0.5 nm down to 0.1 nm, 0.1 nm increments, and then reconstructed the spectrum of the summed response.

As shown in Figure 4c, the Penrose-Moore solution provided a resolution close to 0.5 nm, consistent with the design specifications. However, the LASSO regression improved the resolution close to 0.2 nm, demonstrating its ability to enhance spectral details thanks to its sparse nature[22,28-30].

Accurate amplitude recovery across the spectrum is crucial when comparing complex spectra. Figure 4b illustrates that the amplitude of the laser peak reconstructed by the Penrose-Moore pseudoinverse varied significantly from the power meter's readings. The LASSO regression did not show a clear nonlinear trend, although it exhibited higher noise. As expected, both regression methods failed to estimate the amplitude correctly close to the Littrow and a-Littrow wavelengths. Nonetheless, LASSO regression showed promise for a more robust reconstruction, particularly in handling complex but sparse spectra like those encountered in Raman spectroscopy.

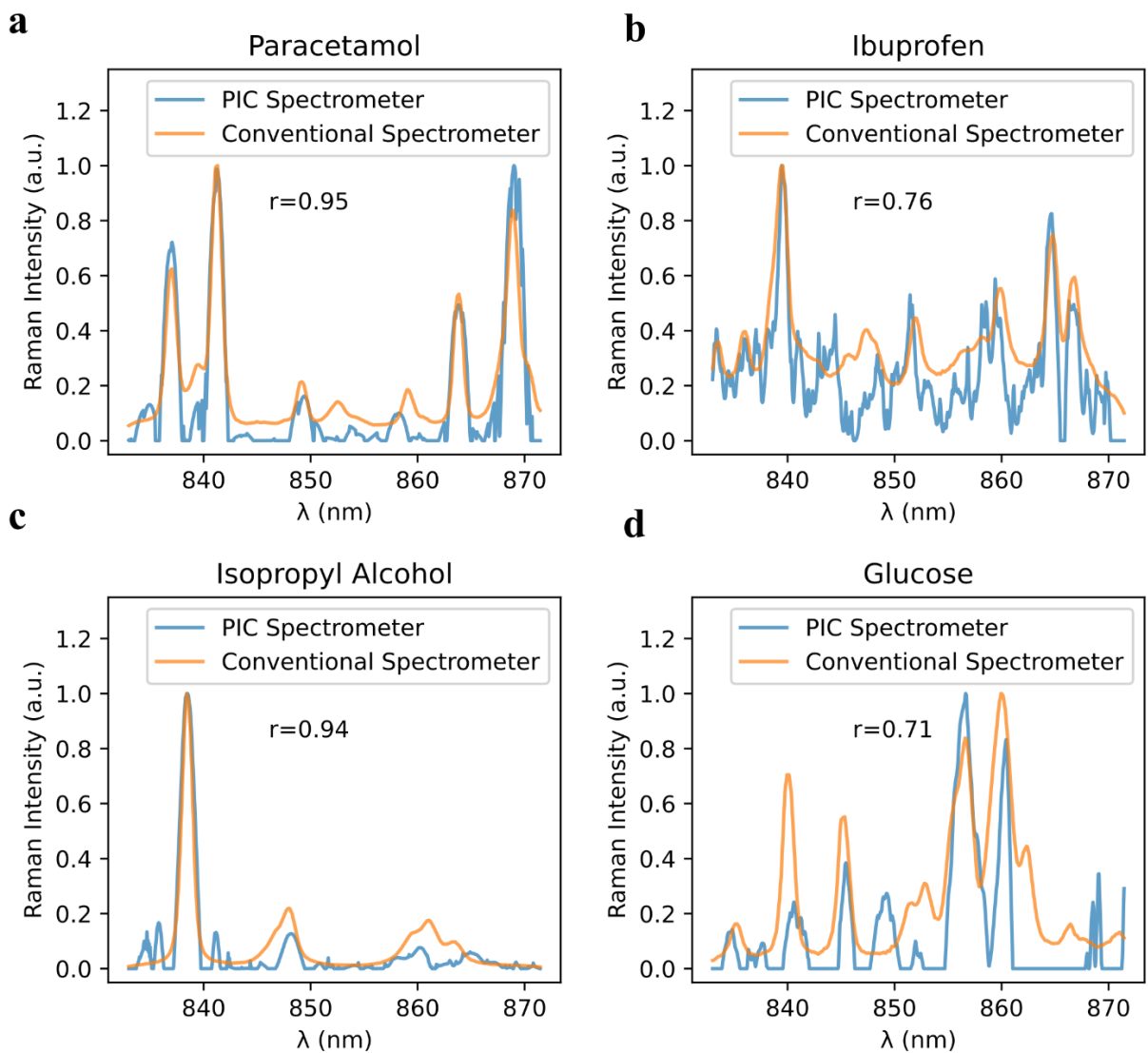

Figure 5. **Comparison of Raman spectra from a conventional spectrometer and the FTS reconstruction for four substances. a.** Paracetamol. **b.** Ibuprofen. **c.** 100% isopropyl alcohol solution. **d.** Pure glucose powder. The Pearson correlation coefficients, indicating the similarity between each reconstructed spectrum and the conventional spectrometer measurements, are shown for each substance.

**Raman Spectrum Reconstruction**

We successfully demonstrated experimental Raman spectrum reconstruction and material identification of four different substances: Paracetamol, Ibuprofen, isopropyl alcohol (IPA), and pure glucose. The experimental setup was similar to the T-matrix measurement configuration, with an additional excitation light at 785 nm, a focusing lens to illuminate the substances and collect the Raman signal, and filters to block the excitation light.

We evaluated Raman signal reconstruction using Penrose-Moore pseudoinverse, LASSO, Ridge, and Elastic-Net regression combined with Savitzky-Golay filtering, both numerically and experimentally. All methods outperformed the pseudoinverse method, consistent with the reports in the literature[22,28–30]. Under low SNR conditions, LASSO and Elastic-Net regression performed similarly, while Ridge produced artificial peaks due to its smoothing effect, making it less suitable for material identification. Since Elastic-Net requires optimizing two hyperparameters versus LASSO's one, we focused on LASSO regression in this article.

Figure 5 represents the Raman spectra obtained from a conventional spectrometer and those reconstructed using our FTS, using LASSO regression with Savitzky-Golay filtering. To quantify the similarity between the conventional and FTS-reconstructed spectra, we used the Pearson correlation coefficient ($r$), a standard technique for assessing spectral similarity[43]. A value of $r > 0.7$ indicates a good match, while $r > 0.9$ signifies a very high correlation. The similarity between the spectra was influenced by factors such as signal strength and spectral complexity[44]. The difference in Raman signal strength between the substances can be because of their Raman cross-section as well as their scattering efficiency i.e. solution, fine particles, or rough surfaces. IPA and Paracetamol showed the highest match, due to the spectral simplicity and higher Raman signal, respectively. Pure glucose exhibited the lowest match, which can be attributed to its rich spectrum with multiple broad peaks, and low Raman signal.

The spectrum reconstruction in Figure 5 used a set of 100 images, each captured with 10 seconds of integration time, for each substance to keep consistency between experiments. For Paracetamol and IPA, a single image was sufficient to reconstruct a spectrum with $r > 0.9$, showing a prospect for the practical use of this FTS for material identification.

In general, lower $r$ were caused by several factors, including the appearance of artificial peaks due to spectral complexity and low optical power, and the disappearance of the baseline because of LASSO regression. Additionally, slight mismatches in peak positions and relative amplitudes were observed. The peak position discrepancies were likely influenced by the smoothing process, while variations in relative peak amplitudes may be due to noise introduced by the regression as shown earlier in Figure 4b.

It should be noted that the Savitzky-Golay filtering decreases the effective resolution, and when multiple complex spectra overlap, a need for higher spectral resolution arises. An FTS design with higher inherent resolution and a T-matrix measurement with smaller steps can be employed to boost the resolution. Alternatively, more complex spectrum reconstruction techniques, such as Elastic-Net, relying on multiple L1-norm, L2-norm, and first derivative minimization coefficients can be used[22,28–30].

We explored concentration detection for IPA solutions, and observed a linear relationship with approximately 90% accuracy down to 25% IPA concentration, though the LASSO regression hyperparameter had a noticeable impact on the results. This suggests a need for further research to improve the signal-to-noise ratio.

The collection efficiency of the FTS must be further improved for an accurate concentration measurement. The current FTS collects a maximum of 3% of the Raman signal in an ideal scenario where the Raman signal is confined within the excitation spot i.e. confocal arrangement and non-scattering opaque medium.

One way of efficiency improvement involves increasing the optical throughput of the optical system. 160 apertures support a throughput that corresponds to the collection of Raman signal from a spot with a length of approximately 79 μm with an NA of 0.55 in one direction. The current optical system collects the signal from an excitation spot diameter of 29 μm with NA= 0.55 considering a non-scattering opaque substance, filling only part of the throughput supported by the chip. In this case, increasing the numerical aperture of the optical system would increase the efficiency by collecting more Raman signal. However, Raman signal is not confined within the excitation spot for a diffusive medium and it can extend several millimeters from the excitation spot[45], showing that the collection optics may not be the limiting factor.

In addition to improvement in the optical system, the optical throughput of our current array is limited by only one single-mode waveguide in the perpendicular direction, allowing collection from a spot with a width of only 0.5 μm. This small collection area significantly limits the efficiency of our spectrometer. To make the optical throughput more symmetrical, one could stack multiple layers of waveguides or employ a two-dimensional array of grating couplers. Another possibility is to use a complex optical component consisting of a multiplexed prism array that transfers the throughput between axes for coupling into the current structure[46].

To reconstruct a spectrum with high SNR, comparable to conventional spectrometers, the optical throughput of the FTS must be increased substantially. Number of apertures in the range of $10^2 - 10^3$ provides a throughput similar to a typical Raman spectrometer as aforementioned. However, the entire spectrum splitting into multiple channels degrades the system SNR due to the shot and thermal noise from multiple channels. Ignoring the photonics component losses, the splitting of the signal creates a need for $10^2 - 10^3$ more optical throughput[44], thus requiring $10^4 - 10^6$ input apertures, to have an SNR comparable to a conventional spectrometer.

Further improvements in reconstructed spectrum quality could also be achieved by optimizing the interferometer and VGC design. Apodized gratings, for instance, could provide better phase matching with the standing wave enhancing the modulation depth and expanding the working range of the interferometers.

## Outlook and Discussion

In this work, we presented an experimental demonstration of Raman spectroscopy for various substances using a highly scalable 150 nm thick SiN waveguide platform with a novel, compact, edge-coupled FTS with 160 input apertures and interferometers. The spectral range and resolution of the FTS were characterized close to the design values as 39.75 nm and 0.51 nm, respectively. We showed that LASSO regression can enhance the resolution down to 0.2 nm while providing a relatively accurate reconstructed peak amplitude compared to the Penrose-Moore pseudoinverse.

The Pearson correlation coefficient reached up to $r = 0.95$, indicating an outstanding match with the Raman spectrum from a conventional spectrometer. We observed that a spectrum with high correlation can be reconstructed down to 10 seconds of integration time using LASSO regression combined with Savitzky-Golay filtering.

To the best of our knowledge, this work represents the first demonstration of an integrated photonics chip spectrometer for Raman spectroscopy. The wafer-scale CMOS-compatible fabrication of the chip allows highly scalable integration. Additionally, this structure does not require any high-resolution imaging, indicating a possibility of direct integration to a photon detector chip and paving the path for compact Raman spectrometers. We believe this research will contribute significantly to the development of wearable devices for non-invasive biomarker detection, affordable hand-held spectrometers, and lightweight instruments for space exploration.

**Acknowledgments**: The authors thank PIC characterization team at Photonic View, Dr. Xiangwei Meng, and Laisheng He for their contributions to the fabrication process, as well as the CUMEC team for their support in developing the SiN PIC platform presented in this article. C.C. is supported by Guangci Innovative Technology Program (KY2023810) and Guangci Talent Program (RC20240018). We also acknowledge Guangci Deep Mind Project of Ruijin Hospital-Shanghai Jiao Tong University School of Medicine.

**Author contribution**: S.K led the design and measurements, compiled and analyzed the spectrometer data, and wrote the manuscript. X.L and Z.D. did the design and layout, helped with the measurements, analyzed the component data, and discussed the core results. Z.Z. envisioned the novel interferometer design and did the first iteration of the design and layout with Z.D. X.Q. and Y.X performed the measurements under the guidance of S.Z., and did the initial analysis. C.C. envisioned the project and roadmap while helping supervise the progress.

---

## Methods

### Device Simulation

The output response of the vertical grating coupler (VGC) and the interferometer with the shortest cavity ($L_0 = 4.555\,\mu m$) were simulated using the finite-difference time-domain (FDTD) method with Lumerical FDTD software. These simulations also provided the effective refractive index ($n_{eff}$). The interferometer and VGC structure were simulated in 2D and 3D for comparison. After confirming similar results through a mesh convergence test, all subsequent optimization simulations were performed in 2D, using a uniform mesh of $5\,nm \times 5\,nm$ around the VGC region and $28\,nm \times 5\,nm$ at the cavity region.

Simulations were conducted under four conditions: (1) with a top-oxide air interface and no VGC reflector, (2) without a top-oxide air interface and with VGC reflector, (3) with both a top-oxide air interface and VGC reflector, (4) without a top-oxide air interface and no VGC reflector. Pre-fabrication simulations assumed a waveguide thickness of $t_{SiN} = 150\,nm$ and a mirror inclination angle of $0°$, while post-fabrication simulations were adjusted to match the estimated waveguide thickness and mirror inclination, $t_{SiN} = 136\,nm$ and $6°$. For longer interferometers, due to the extended simulation times, a numerical model was developed instead of relying on direct FDTD simulations.

### Matching the Numerical Model with Measurement Results

The effective refractive index ($n_{eff}$) for the Littrow wavelength and the center wavelength between two Littrow wavelengths, a-Littrow, (both denoted as $\lambda_l$) was first calculated from the measurement using the equation $n_{eff}^l = \frac{L\lambda_l}{2\Delta L}$ where $l$ is a positive integer, and $n_{eff}^l$ is the effective refractive index at those wavelengths. Since the T-matrix measurement covered a Littrow and an a-Littrow wavelength, we assumed a linear relationship between $n_{eff}$ and wavelength, modeled as $n_{eff} = m^l \cdot \lambda + n_g^l$, where the slope $m^l$ and group index $n_g^l$ were estimated based on the effective refractive index at the Littrow and a-Littrow wavelengths. The resulting $n_{eff}$ profile showed good agreement with that of a waveguide with a thickness of $t_{SiN} = \sim 136\,nm$, as determined by Lumerical FDTD solver.

The mirror inclination of $6°$ was accounted for by retrieving the reflection amplitude coefficient through FDTD simulations. Several fabrication-induced parameter variations were considered, including the cavity length (due to mask misalignment), effective refractive index (where small variations can cause significant mismatches in longer interferometers), top-oxide thickness $t_{TOX}$, and VGC reflector distance $h$. At each step, the Pearson correlation coefficient between the measurement results and the numerical model for a set of cavity lengths was used as the figure of merit for matching.

The matching procedure began with a sweep of the misalignment correction factor, ranging from -200 nm to 200 nm. Since the shortest cavity responses are most sensitive to length variations, the Pearson correlation coefficients for the five shortest cavities were averaged to find the best match. After identifying the optimal misalignment correction length, the effective refractive index was further optimized for longer interferometers with cavity lengths of 384.565 nm, 288.965 nm, and 193.365 nm. The optimization was performed using a genetic algorithm (from PyGad toolbox[47]) with 50 solutions per population, 10 parents per generation, 5 survivors, and 1000 generations with a saturation stop condition of 50 generations. The parameter search space was defined as $m = (1 \pm 0.001)m^l$ and $n_g = (1 \pm 0.001)n_g^l$.

Once the effective refractive index was optimized, we performed a genetic algorithm search for the top-oxide thickness and VGC reflector distance using the same genetic algorithm settings. The Pearson correlation coefficient was averaged over interferometers with cavity lengths of 384.565 nm, 193.365 nm, 26.065 nm, and 4.555 nm. The search range for the top-oxide thickness ($t_{TOX}$) was from 2992.5 nm to 3657.5 nm, and for the VGC reflector distance ($h$), it ranged from 292.5 nm to 357.5 nm.

### Wafer Fabrication

A 200 mm SiN PIC wafer was fabricated in United Microelectronics Center (CUMEC). Initially, a $SiO_2$ layer was deposited on a 725 µm thick Si wafer using plasma-enhanced chemical vapor deposition (PECVD), then thinned to 2 µm by chemical mechanical polishing (CMP). A 150 nm thick SiN waveguide layer was deposited via low-pressure chemical vapor deposition (LPCVD) and patterned using two masks. The waveguide layer was dry-etched fully and to the depth of 50 nm to define the designed structures.

Next, a 325 nm $SiO_2$ layer was deposited (PECVD followed by CMP), followed by the deposition of a 200 nm Al layer to form the VGC reflector. A further 3 µm of $SiO_2$ (PECVD and CMP) was deposited. The device wafer was then flipped and bonded to a carrier wafer with 1.0 µm of SiO2 (PECVD) using oxide-oxide fusion bonding. The Si substrate of the device wafer was completely removed by grinding and CMP, leaving 1.5 µm of $SiO_2$ on top.

The interferometer mirror was fabricated by dry-etching 4.1 µm of $SiO_2$ and 150 nm of SiN, followed by the deposition of 500 nm Al layer. Finally, an additional $SiO_2$ layer was deposited to bring the total top oxide thickness to 3.325 µm.

### Transform Matrix Measurement

A tunable laser with laser power varying from 21.3

mW to 24.8 mW, depending on the wavelength, covering 805-880 nm range, was used to scan the interferometer responses with 0.05 nm steps. To prevent speckles in the spatial profile, the laser light was directed into an integrating sphere and then guided into free space through a short section of a multimode fiber with a 600 μm core diameter. The light was quasi-collimated by an aspherical lens (L1, f=10 mm, NA=0.55) and split by a beam splitter.

One beam path was imaged onto a camera via two mirrors and a lens (L5, f = 40 mm) to monitor the alignment of the fiber input. The second beam path included two cylindrical lenses (CL1, $f_y$ = 50 mm, and CL4, $f_y$ = 40 mm) arranged in a 2f configuration, which focused and collimated the beam along the y-axis (perpendicular to the chip surface) to relay the beam and enhance the Raman spectroscopy measurements. Finally, the beam was focused along the y-axis using another cylindrical lens (CL5, $f_y$ = 10 mm).

In the axis parallel to the chip edge, two cylindrical lenses (CL2, $f_x$ = 50 mm, and CL3, $f_x$ = 10 mm) in a 2f configuration demagnified the beam by 5x. This created a light sheet with FWHM of 301 μm (width) and 2596 μm (length) for edge coupling into the chip. The chip's output was captured by a 5x objective (O1, NA = 0.14) and imaged onto a Hamamatsu camera using a tube lens (L2, f = 200 mm) and an additional lens (L3, f = 150 mm). The camera recorded a 16-bit image for each wavelength with an integration time of 10 ms and the same procedure was followed for the laser spectrum reconstruction measurements. The output power of each interferometer was obtained by summing the pixel values over a 5x5 pixel area. To mitigate the high-frequency noise in the interferometer responses, a low-pass filter, with a cut-off at the edge of the response peaks in the Fourier domain, was applied.

**Raman Spectroscopy Measurements**

For the Raman spectroscopy measurements, the portion of the T-matrix optical setup that included five cylindrical lenses for forming a light sheet at the chip edge was retained. A 785 nm laser beam with 537 mW power, delivered through a multimode fiber (core diameter 105 μm, NA = 0.22), was quasi-collimated by a lens (L4, f = 30 mm) and filtered with a laser clean-up filter (F1, OD > 5). The collimated beam was then directed via two mirrors and a dichroic mirror and focused onto the substance using a lens (L1, f = 10 mm, NA = 0.55), creating a spot with a 14 μm diameter.

The scattered light, including the Raman signal, was collected and collimated by the same lens. A long-pass edge filter (F1, 785 nm cut off, OD > 6) was used to block the excitation light. The remaining beam was further filtered by a band-pass filter (F2, 850 nm center wavelength, 40 nm full-width half-maximum) to match the Fourier transform spectrometer's range and eliminate any remaining excitation light.

To measure the Raman spectrum using a conventional Raman spectrometer, an additional beam splitter before the cylindrical lenses directed half of the Raman signal through a lens (L5, f = 30 mm) and into a multimode fiber (core diameter 105 μm, NA = 0.22). The beamsplitter was removed during the chip spectrometer measurements to maximize signal power. The Raman signal was focused on the chip edge via the cylindrical lenses, and the output was recorded using a Hamamatsu camera. For each measurement, 100 16-bit images with 10 s of integration times were recorded. Before each Raman measurement, background images were captured by turning the laser off. For processing, the sum of background images was subtracted from the sum of the Raman measurement images. The interferometer output power was then determined by summing the pixel values within a 5x5 pixels region.

All substances were placed in a glass cuvette. Both the Ibuprofen in a sustained capsule (Fenbid®) and Paracetamol (Jinzhou Jiutai Pharmaceutical) pill were crushed into fine particles. Isopropyl alcohol (2-Propanol) and Glucose (D-(+)-Glucose, ⩾99.5% (GC)) were sourced from Sigma Aldrich.

**Spectrum Reconstruction**

The interferometer values were first normalized by a Gaussian profile to account for the light sheet's intensity distribution. Laser and Raman spectra were reconstructed from the FTS response using different regression techniques, including Penrose-Moore pseudoinverse, LASSO regression, Ridge regression, and Elastic-Net regression[48].

Data processing was performed in Python 3.11. Penrose-Moore pseudoinverse was calculated using Numpy library, while the other regression techniques were implemented via scikit-learn[49]. Savitzky-Golay smoothing was applied using the Scipy library to improve the quality of the reconstructed spectrum.

Each regression method aimed to solve the equation $Ax = y$ for the unknown spectrum vector $x$, where $A$ is the 2D transform matrix (varying by interferometer and wavelength), and $y$ represents the measured output power from the FTS. The challenge of finding an optimum solution arises because $A$ is a wide rectangular matrix (160 × 1501), with a rank equal to the number of rows, making the system underconstrained with infinite possible solutions[50].

The Least squares solution ($\bar{x} = A^+ y$) was calculated using the Penrose-Moore pseudoinverse, $A^+$, obtained through singular value decomposition (SVD) of $A = U\Sigma V^T$, where $U$ and $V$ are unitary matrices, and $\Sigma$ contains the square roots of the eigenvalues.

LASSO regression was used to impose sparsity and limit the coefficient magnitude, especially useful in low SNR conditions, by minimizing the sum of squared errors with an additional L1-norm penalty:

$$\min_x \ \{\|y - A\bar{x}\|_2^2 + \lambda^{LASSO} \|\bar{x}\|_1\} \qquad (1)$$

where the first term represents the least square regression term and the second term is an extra penalty on the L1-norm scaled with a hyperparameter of $\lambda^{LASSO}$.

Ridge regression (Tikhonov regularization) minimizes the spectral coefficients without driving them to zero by applying a penalty on the L2-norm with a hyperparameter $\lambda^{RIDGE}$:

$$\min_{x} \; \{\|y - A\bar{x}\|_2^2 + \lambda^{RIDGE}\|\bar{x}\|_2\} \qquad (2)$$

Elastic-Net regression combines both L1 and L2 penalties for smoothness and coefficient reduction:

$$\min_{x} \; \{\|y - A\bar{x}\|_2^2 + \lambda_1^{EN}\|\bar{x}\|_1 + \lambda_2^{EN}\|\bar{x}\|_2\} \qquad (3)$$

All regression methods (except the Penrose-Moore pseudoinverse) were influenced by their hyperparameters and the quality of spectrum reconstruction was further enhanced through Savitzky-Golay smoothing, whose window length and polynomial order were also optimized.

Optimization of all parameters, including regression hyperparameters and smoothing factors, was performed using a genetic algorithm (PyGad library[47]). The objective was to maximize the Pearson correlation coefficient between the spectrum obtained from a conventional spectrometer and the FTS-reconstructed spectrum. Each genetic algorithm ran for 100 generations with 50 members per generation and 25 parents, keeping one survivor per generation. The optimum hyperparameters $\lambda^{LASSO}$, $\lambda^{RIDGE}$ and $\lambda_1^{EN}$ were searched in a logarithmic range from $10^{-12}$ to $10^{8}$, due to the varying signal strength of the substances. The L1 to L2 hyperparameter ratio $\lambda_1^{EN} \big/ \lambda_2^{EN}$ of Elastic-Net regression was optimized between 0 and 1, while the Savitzky-Golay smoothing window length was searched between 10 to 150, and polynomial orders between 3 and 8.